\documentclass[12pt]{article}

\usepackage{graphicx}
\usepackage{caption}
\usepackage{subcaption}
\usepackage{mathptmx}           
\usepackage[applemac]{inputenc} 
\usepackage{dsfont}
\usepackage{clrscode}             
\usepackage{url} 
\usepackage{color}
\usepackage[usenames,dvipsnames]{xcolor} 
\usepackage{ marvosym }
\usepackage{enumitem}
\usepackage{natbib}
\usepackage{hyperref}

\begin{document}
\sloppy

\title{\bf A Java Implementation of Parameter-less Evolutionary Algorithms}

\author{   {\bf José C. Pereira}\\
            \small CENSE and DEEI-FCT\\
            \small Universidade do Algarve\\
            \small Campus de Gambelas\\
            \small 8005-139 Faro, Portugal\\
            \small unidadeimaginaria@gmail.com
\and
		 {\bf Fernando G. Lobo}\\
            \small CENSE and DEEI-FCT\\
            \small Universidade do Algarve\\
            \small Campus de Gambelas\\
            \small 8005-139 Faro, Portugal\\
            \small fernando.lobo@gmail.com         
}
\date{}
\maketitle

\begin{abstract}
The Parameter-less Genetic Algorithm was first presented by Harik and Lobo in 1999 as an alternative to the usual trial-and-error method of finding, for each given problem, an acceptable set-up of the parameter values of the genetic algorithm. Since then, the same strategy has been successfully applied to create parameter-less versions of other population-based search algorithms such as the Extended Compact Genetic Algorithm and the Hierarchical Bayesian Optimization Algorithm. This report describes a Java implementation, \emph{Parameter-less Evolutionary Algorithm} (P-EAJava), that integrates several parameter-less evolutionary algorithms into a single platform.
Along with a brief description of P-EAJava, we also provide detailed instructions on how to use it, how to implement new problems, and how to generate new parameter-less versions of evolutionary algorithms.  

At present time, P-EAJava already includes parameter-less versions of the Simple Genetic Algorithm, the Extended Compact Genetic Algorithm, the Univariate Marginal Distribution Algorithm, and the Hierarchical Bayesian Optimization Algorithm. The source and binary files of the Java implementation of P-EAJava are available for free download at \textcolor{Blue}{\href{https://github.com/JoseCPereira/2015ParameterlessEvolutionaryAlgorithmsJava}{https://github.com/JoseCPereira/2015ParameterlessEvolutionaryAlgorithmsJava}}.
 
\end{abstract}


\section{Introduction}\label{sec:intro}
The parameter-less genetic algorithm (P-GA) was first presented by \cite{HarikLobo:99} as an alternative to the usual trial-and-error method of finding, for each given problem, an acceptable set-up of the parameter values of the genetic algorithm. 

Shortly after, \cite{PelikanLobo:99} showed that, even in the worst-case scenario, the total number of function evaluations performed in one run of the P-GA does not increase significantly
with respect to the number of function evaluations performed by the standalone GA tuned with an optimal fixed population size. Moreover, the authors pointed out that the worst-case scenario was extremely improbable to occur and that the expected performance of the P-GA in practice should be much more benevolent. Therefore, \cite{PelikanLobo:99} concluded that the P-GA is an efficient method of eliminating the parameters of the simple GA .

Since then, the same strategy has been successfully applied to develop other parameter-less evolutionary algorithms such as the Parameter-less Extended Compact Genetic Algorithm (P-ECGA)  \citep{Lobo:00}, the Parameter-less Genetic Programming Algorithm (P-GPA) \citep{SpinosaPozo:02}, and the Parameter-less Hierarchical Bayesian Optimization Algorithm (P-HBOA) \citep{PelikanHartLin:07}. Building on these results, we implemented in Java the \emph{Parameter-less Evolutionary Algorithm} (P-EAJava) that integrates several parameter-less evolutionary algorithms into a single platform.

At present time, P-EAJava already includes the P-GA, the P-ECGA, the P-HBOA, and a parameter-less version of the Univariate Marginal Distribution Algorithm \citep{Muhlenbein:96}. The source and binary files of the Java implementation of P-EAJava are available for free download at \textcolor{Blue}{\href{https://github.com/JoseCPereira/2015ParameterlessEvolutionaryAlgorithmsJava}{https://github.com/JoseCPereira/2015ParameterlessEvolutionaryAlgorithmsJava}}.

The P-EAJava uses as a common working base the standard versions of the Simple Genetic Algorithm \citep{Holland:75,Goldberg:89a,EibenSmith:03}, the Univariate Marginal Distribution Algorithm \citep{Muhlenbein:96}, the Extended Compact Genetic Algorithm \citep{Harik:99a}, and the Hierarchical Bayesian Optimization Algorithm \citep{PelikanGoldberg:06}. The Java implementation of these four standard EAs is presented in detail in another arXiv report from the same authors. The corresponding source code is also available for free download at \textcolor{Blue}{\href{https://github.com/JoseCPereira/2015EvolutionaryAlgorithmsJava}{https://github.com/JoseCPereira/2015EvolutionaryAlgorithmsJava}}.

The remainder of this paper is organized as follows. In Section \ref{sec:PGA} we briefly describe the main concepts of the P-GA that are the basis for the P-EAJava implementation. In Section \ref{sec:P-EAJava} we discuss the P-EAJava implementation itself and provide detailed instructions on how to use it, how to implement new problems with it, and how to generate new parameter-less versions of evolutionary algorithms. 


\section{The Parameter-less Genetic Algorithm}\label{sec:PGA}
The P-GA establishes an evolutionary race among an unbounded number of independent populations. All populations are evolved using the same simple GA which has most of its parameters set a priori to robust fixed values, in accordance with facetwise theoretical studies \citep{Goldberg:1993,Thierens:93}. The only exception is the population size which is dynamically adapted using a population sizing method.

In the P-GA each independent population is uniquely identified by its size, $N_i$. This size doubles between consecutive populations, according to the identity
\begin{equation}\label{eq:N}
N_i = 2^i\,N_0 \,, \quad \forall i \in \mathds{N} \enspace .
\end{equation}
where $N_0$ is the initial population size, chosen to be a small enough value such as $N_0=10$.

\begin{figure}[t!]
\centering
\includegraphics[width=0.9\textwidth]{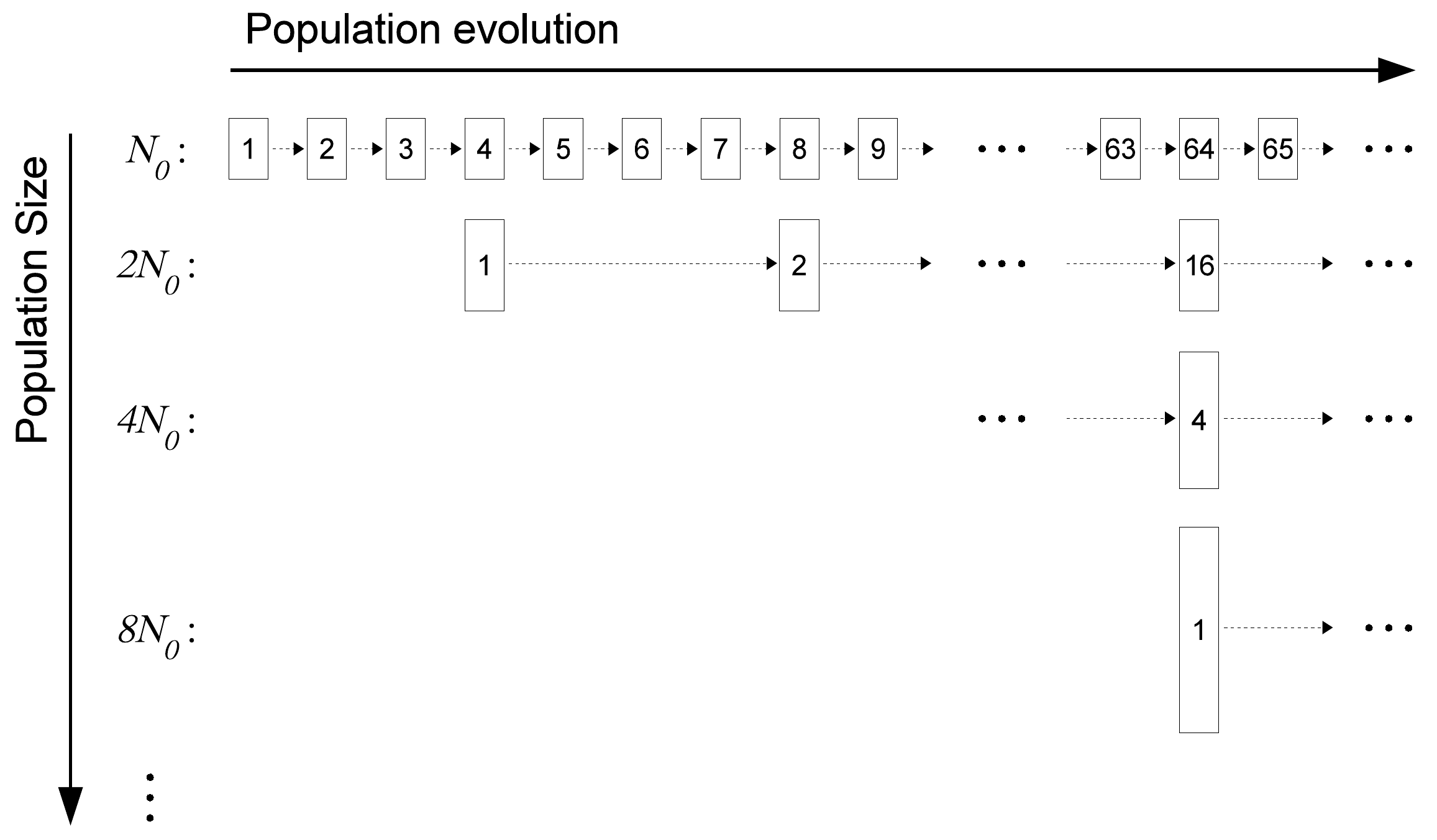}
\caption{The P-GA establishes a race among an unbounded number of populations with exponentially increasing size that are evolved using a simple GA. For each 4 generations performed with population $2^iN_0$ there will be one generation performed with population $2^{i+1}N_0$. The initial population size $N_0$ is chosen to be a small enough value such as $N_0=10$.}
\label{fig:P-GA}
\end{figure}

The P-GA starts by performing 4 generations of population $N_0$, after which it evolves population $N_1$ for a single generation. This process is then repeated so that, for each 4 generations performed with population $N_0$, there will be one generation performed with population $N_1$. The same evolutionary procedure  is used recursively for all other populations so that, throughout the entire run of the algorithm, for each 4 generations performed with population $N_i$, there will be one generation performed with population $N_{i+1}$. Figure \ref{fig:P-GA} depicts a graphical representation of a possible instance of this population evolution.

The P-GA works under the assumption that solution quality grows monotonously with the population size \citep{LoboLima:07}. In addition, the algorithm also assumes that this improvement in solution quality happens at a decreasing rate, i.e., when the same increase in population size happens between larger and larger populations, the gains in solution quality are expected to be fewer and fewer. For these reasons, the evolutionary race is established between populations with \emph{exponentially increasing} size. Simultaneously, the P-GA allows smaller populations to use more computational resources, because if two different populations produce solutions with similar quality then the smaller population should be preferred, all other things being equal.

\begin{figure}[t!]
\centering
\includegraphics[width=0.65\textwidth]{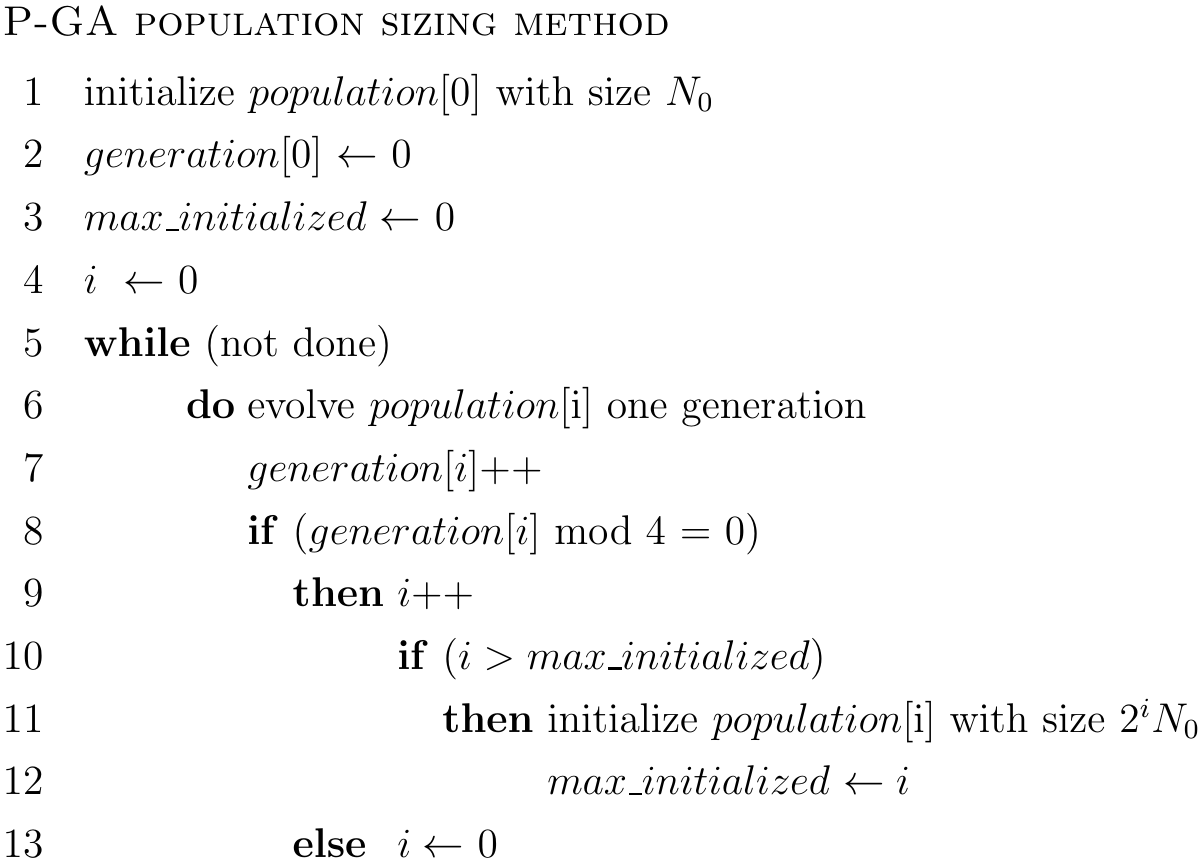}
\caption{Pseudocode for the population sizing method employed by the P-GA.}
\label{fig:PGAPseudo1}
\end{figure}

In their paper, \cite{HarikLobo:99} used a counter of base 4 to implement the population sizing method of the P-GA (see also, \cite{Lobo:00, PelikanLobo:99} for detailed descriptions of this counter). However, \cite{PelikanHartLin:07} suggested a somewhat simpler implementation for that sizing method and used it to propose a parameter-less version of the HBOA. The same simpler implementation was adopted to develop the work reported in this paper. Figure \ref{fig:PGAPseudo1} depicts the pseudocode for the population sizing method as presented by Pelikan et al, where only the name in the title had to be changed to make it work for the simple genetic algorithm.

In the P-GA, the performance of a population at any point in time is measured by the current average fitness of all its individuals and by how many resources it has consumed, i.e., by how many fitness function evaluations it has already performed. Naturally, a population that has consumed more resources is expected to have a better fitness than any other population that has performed less fitness evaluations. When this is not the case, the more costly populations should be discarded. In practice, whenever population $N_i$ obtains an average fitness at least as good as the average fitness of population $N_{i-1}$, the P-GA eliminates all smaller populations $N_k$, with $k < i$. Note that, strictly speaking, the P-GA could maintain active those smaller populations that eventually still have a better fitness than $N_i$. However, if $N_i$ has already caught up with $N_{i-1}$, it will definitely catch up with all other smaller populations, rather sooner than later. Therefore, it is more efficient not to waste any more time with these populations and focus all resources on those populations that promise to deliver better quality solutions. 

For the sake of simplicity the P-GA employs a genetic algorithm without a mutation operator. In the absence of mutation, all individuals in a population eventually become identical. In such a state, the population is said to have converged and can no longer generate any new individuals and will not improve its average fitness. In this case, the P-GA stores the information concerning the best solution found and eliminates not only the converged population, but all other smaller populations. This is efficient and effective, because once a population has converged, all smaller populations will also converge in the near future, and since solution quality increases along side with population size, there is nothing to gain by waiting for them to do so.

The P-GA is designed to run forever, physical constraints aside, because in practice the quality of the optimal solution is often unknown for many problems, making it impossible to distinguish, for instance, when the algorithm has reached the optimum result from when it simply got ``stuck'' in some plateau of the search space. Therefore, the P-GA leaves to the user the decision when to stop the computation, based on the quality of the solutions already found and on the time and resources that she or he is willing to spend. 

With an ever increasing set of larger and larger populations, the P-GA would soon exhaust all computational resources available. By discarding populations that either do not comply with the parameter-less invariant or can no longer improve the quality of its solutions, the P-GA ensures that the actual number of active populations, although unbounded, will not grow excessively large with time. In fact, this discarding process, which is more intensive for smaller populations, is one of the main factors behind the relatively small amount of extra computational resources that the P-GA requires in relation to the standalone GA working with an optimal fixed population size. 

\

The implementation of a parameter-less version of the HBOA by \cite{PelikanHartLin:07} showed in practice that the parameter-less strategy proposed by \cite{HarikLobo:99} was not tied to a specific evolutionary algorithm. The same result had also been shown by \cite{Lobo:00} with the implementation of the P-ECGA which was used by the author to solve an electrical network expansion problem. Likewise, \cite{PelikanHartLin:07} used the P-HBOA to successfully solve artificial hierarchical and nearly decomposable problems, and a 2D version of the real world Ising Spin Glasses problem.
In the next section, we describe how to use the P-EAJava.


\section{The Parameter-less Evolutionary Algorithm}\label{sec:P-EAJava}

The Parameter-less Evolutionary Algorithm (P-EAJava) is a Java implementation of the population sizing method employed by the P-GA (see Figure \ref{fig:PGAPseudo1}). However, by decoupling the population sizing method from the evolutionary algorithm itself, the P-EAJava allows the creation of parameter-less versions of other evolutionary algorithms in a straightforward manner. 

Naturally, the P-EAJava works with algorithms for which it is reasonable to presume the same assumptions as the simple GA, namely that solution quality grows monotonously with the population size, but at a decreasing rate and that it is possible to effectively automate the adaptation process of all the algorithm's parameters, with the possible exception of the population size. 

At present time, an evolutionary algorithm must satisfy the following set of constraints in order to be integrated in the P-EAJava:

\begin{enumerate}
\item The algorithm represents possible solutions (individuals) as strings of zeros and ones.
\item All individuals have the same string size.
\item The population size remains constant throughout a complete run of the algorithm.
\end{enumerate}
 
 Naturally, it should be possible to further generalize the P-EAJava and eliminate or at least weaken such constraints.

  \subsection{How to use the P-EAJava}
 
 The P-EA is a Java application developed with the Eclipse\footnote{Version: Kepler Service Release 2} IDE. The available code is already compiled and can be executed using the command line.

 \paragraph{Run the P-EAJava from a command line}
 
\begin{enumerate}
\item Unzip the source file \textit{2015ParameterlessEAs.zip} to any directory.
\item Open your favourite terminal and execute the command\\ 
\vspace{-.5cm}

{\tt{~~cd [yourDirectory]/2015ParameterlessEvolutionaryAlgorithmsJava/bin}}\\
\vspace{-.3cm}

where {\tt{[yourDirectory]}} is the name of the directory chosen in step 1.
 
\item Execute the command\\ 
\vspace{-.5cm}

{\tt{~~java com/z\_PEA/PARAMETERLESS ./PEAParameters.txt}}
\end{enumerate}

\vspace{.3cm}
The argument ``PEAParameters.txt'' is in fact the name of the file containing all the options concerning the P-EA settings and can be changed at will. 

After each execution of a single or multiple runs, the P-EAJava produces one output file --\emph{PARAMETERLESS\_*\_*.txt} -- that records how each run progressed in terms of population size, best  current fitness, among other relevant information. Additionally, the P-EAJava also creates the file -- \emph{PARAMETERLESS-STATS\_*\_*.txt} -- that stores some of the statistics  necessary for analyzing the behaviour of the parameter-less algorithms over multiple runs.

\

At present time, the P-EAJava version made available with this paper already includes four parameter-less algorithms: 
\begin{itemize}
\item Parameter-less Genetic Algorithm,
\item Parameter-less Univariate Marginal Distribution Algorithm,
\item Parameter-less Extended Compact Genetic Algorithm, 
\item Parameter-less Hierarchical Bayesian Optimization Algorithm
\end{itemize}

The current code also includes a set of test problems that can be solved using some or all of the previous algorithms. Here is the problem menu:\\

\textit{ZERO} Problems \textit{\qquad \qquad \qquad \qquad \qquad ONE} Problems
\vspace{.4cm} 

\begin{tabular}{rlrl}
~~0 $\rightarrow$ & ZeroMax & ~~10 $\rightarrow$ & OneMax\\
~~1 $\rightarrow$ & Zero Quadratic & ~~11 $\rightarrow$ & Quadratic\\
~~2 $\rightarrow$ & Zero 3-Deceptive & ~~12 $\rightarrow$ & 3-Deceptive\\
~~3 $\rightarrow$ & Zero 3-Deceptive Bipolar & ~~13 $\rightarrow$ & 3-Deceptive Bipolar\\
~~4 $\rightarrow$ &  Zero 3-Deceptive Overlapping & ~~14 $\rightarrow$ & 3-Deceptive Overlapping\\
~~5 $\rightarrow$ & Zero Concatenated Trap-k & ~~15 $\rightarrow$ & Concatenated Trap-k\\
~~6 $\rightarrow$ & Zero Uniform 6-Blocks & ~~16 $\rightarrow$ & Uniform 6-Blocks
\end{tabular}\\ 

\

\textit{Hierarchical} Problems	
\vspace{.4cm} 

\begin{tabular}{rl}
~~21 $\rightarrow$ &  Hierarchical Trap	One\\
~~22 $\rightarrow$ & Hierarchical Trap	Two
\end{tabular}\\

\

The Zero problems always have the string with all zeros as their best individual. The One problems are the same as the Zero problems but their best individual is now the string with all ones. A description of these problems can be found, for instance, in \cite{PelikanPazGoldberg:2000}. The Hierarchical problems are thoroughly described in \cite{Pelikan:05}. 

It is also possible to define a noisy version for any of the previous problems. This is done by adding a non-zero Gaussian noise term to the fitness function.

The source code that implements all the problems mentioned in this section can be found in the file  \textit{src/com/z\_PEA/Problem.java}.

As mentioned previously, all options concerning the parameter-less strategy are in the file \textit{PEAParameters.txt}. In particular, it is in this file that are made the choices for the problem to be solved and for the parameter-less algorithm that is going to solve the chosen problem. 

To choose a particular problem the user must set the value of the following three options:\\

\begin{tabular}{rl}
~~Line 81:  &  \textit{problemType}\\
~~Line 90:  & \textit{stringSize}\\
~~Line 107: & \textit{sigmaK} ~~~~~(defines the noise component)
\end{tabular}\\

Analogously, to choose a particular parameter-less algorithm it is necessary to set the values of the following two options:\\

\begin{tabular}{rl}
~~~~~~~Line 125:  &  \textit{eAlg}\\
~~~~~~~Line 135:  & \textit{eaParamFile} ~~~~~(defines the name of the corresponding parameters file)
\end{tabular}\\

All other options are set to default values and their role in the parameter-less strategy is explained with detail in the file's comments. This is also true for the parameters specific to each of the implemented algorithms which are defined in four separate files:\\

\begin{tabular}{rl}
~~SGA:  &SGAParameters.txt\\
~~UMDA:  &UMDAParameters.txt\\
~~ECGA:  &ECGAParameters.txt\\
~~HBOA:  &HBOAParameters.txt\\	
\end{tabular}\\

Note that the default settings defined in these three files were chosen to ensure a robust behavior of the corresponding algorithms, in accordance with current theory. Therefore, the user is advised to proceed with caution when performing any changes in those settings. In fact, the whole idea behind the parameter-less strategy is to eliminate the need of such fine tuning when solving a particular problem. After choosing a problem to be solved and a particular algorithm to solve it, the user has only to press the start button and wait until the P-EAJava finds a solution with good enough quality.

\begin{figure}[t!]
\centering
\includegraphics[width=0.85\textwidth]{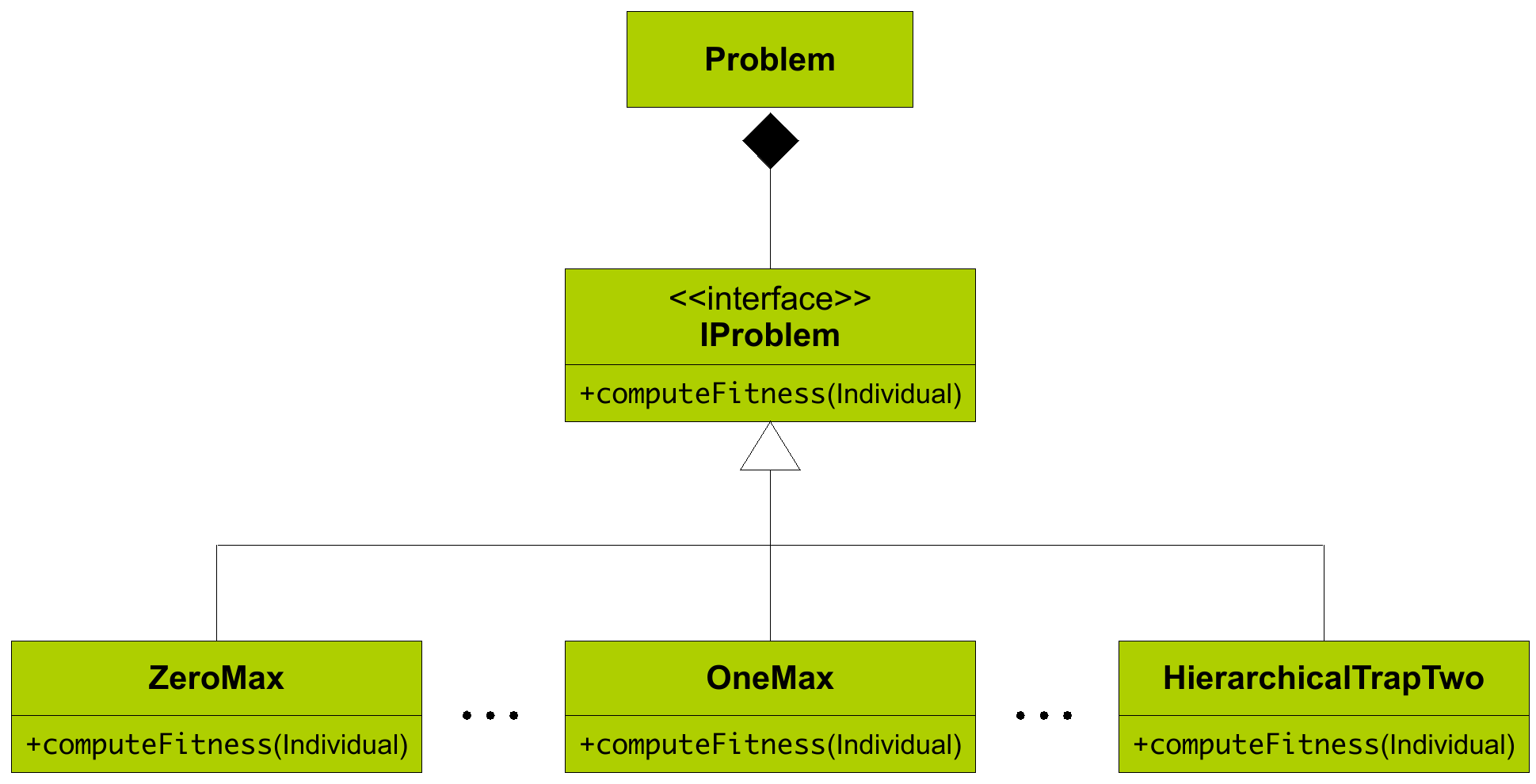}
\caption{The P-EAJava uses the design pattern strategy \citep{Gamma:95} to allow an easy implementation of new problems to be solved by the framework.}
\label{fig:IProblem}
\end{figure}

\subsection{How to implement a new problem with P-EAJava}

The P-EAJava uses the design pattern strategy \citep{Gamma:95} to decouple the implementation of a particular problem from the remaining parameter-less strategy (see Figure \ref{fig:IProblem}). As a consequence, to plug in a new problem to the framework it is only necessary to define one class that implements the interface IProblem and change some input options to include the new choice. The interface IProblem can be found in the file \textit{src/com/z\_PEA/Problem.java}.

In the following let us consider that we want to solve a new problem called \textit{NewProblem} with one of the parameter-less algorithms. To plug in this problem it is necessary to:

\begin{enumerate}
\item Define a class called \textit{NewProblem} in the file \textit{src/com/z\_PEA/Problem.java}. The signature of the class will be
\begin{figure}[h!]
\centering
\includegraphics[width=0.65\textwidth]{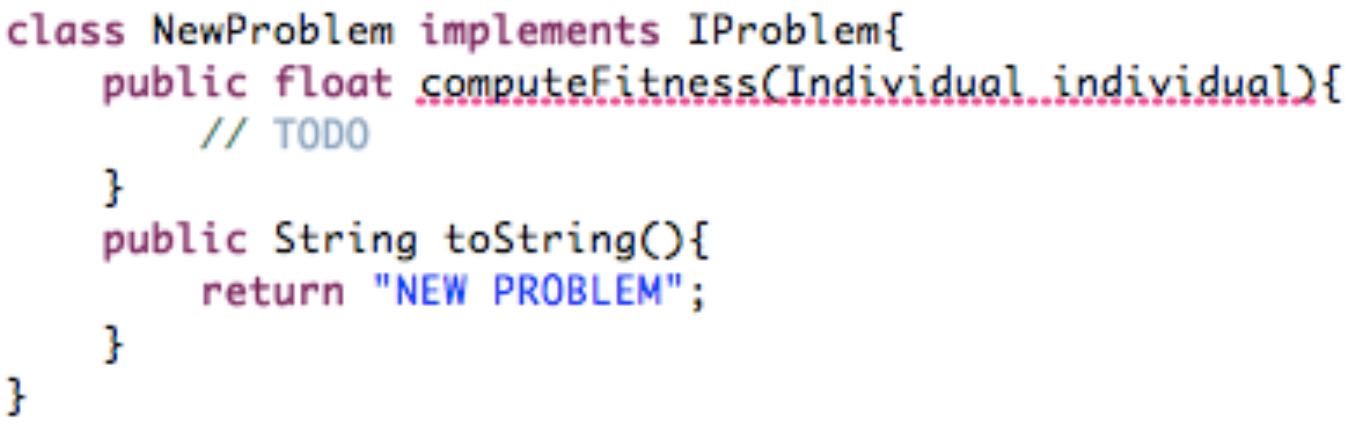}
\end{figure}
\vspace{-.5cm}
\item Code the body of the function computeFitness(Individual) according to the nature of problem \textit{newProblem}. The class \textit{Individual} provides all the necessary functionalities to operate with the string of zeros and ones that represents an individual  (e.g., getAllele(int)). This class can be found in the file \textit{src/com/z\_PEA/Individual.java}.

\item To define the new problem option, add the line

\includegraphics[width=.95\textwidth]{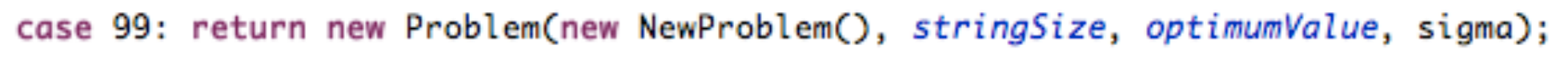}
 to the \textbf{switch} command in line 209 of the file \textit{src/com/z\_PEA/ParParameter.java}. The case number -- 99 -- is a mere identifier of the new problem option. The user is free to choose other value for this purpose. The rest of the line is to be written verbatim. 
 
\item Validate the new problem option value -- 99 -- by adding the case \textit{problemType} == 99 to the conditional in line 105 of the same \textit{ParParameter.java} file.

\end{enumerate}

\vspace{.2cm}
Although not strictly necessary, it is also advisable to keep updated the problem menu in the file \textit{PEAParameters.txt}.

\

\subsection{How to create a new parameter-less algorithm with P-EAJava}

\begin{figure}[t!]
\centering
\includegraphics[width=0.8\textwidth]{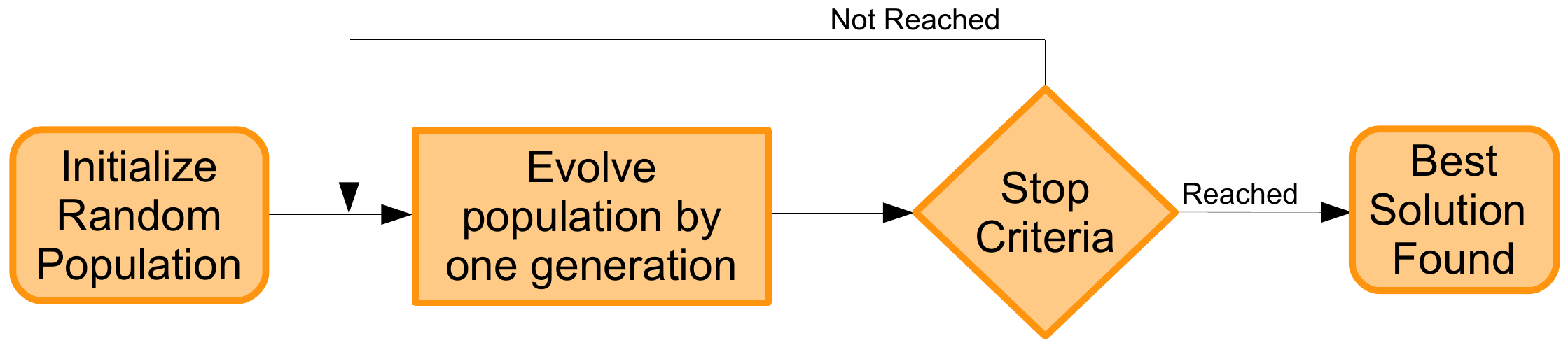}
\caption{General flowchart of a population-based search algorithm. The ParFrame can only integrate algorithms that follow this workflow.}
\label{fig:EAFlow}
\vspace{-.2cm}
\end{figure}

At present time, the P-EAJava is general enough to work with population-based algorithms that follow the workflow depicted in Figure \ref{fig:EAFlow}. As such, the framework already provides all the necessary classes that define the working notions of Population and Stop Criteria. The evolutionary algorithm has only to provide its own classes related with the operators responsible for the evolution of the population in each generation.

\begin{figure}[t!]
\centering
\includegraphics[width=0.7\textwidth]{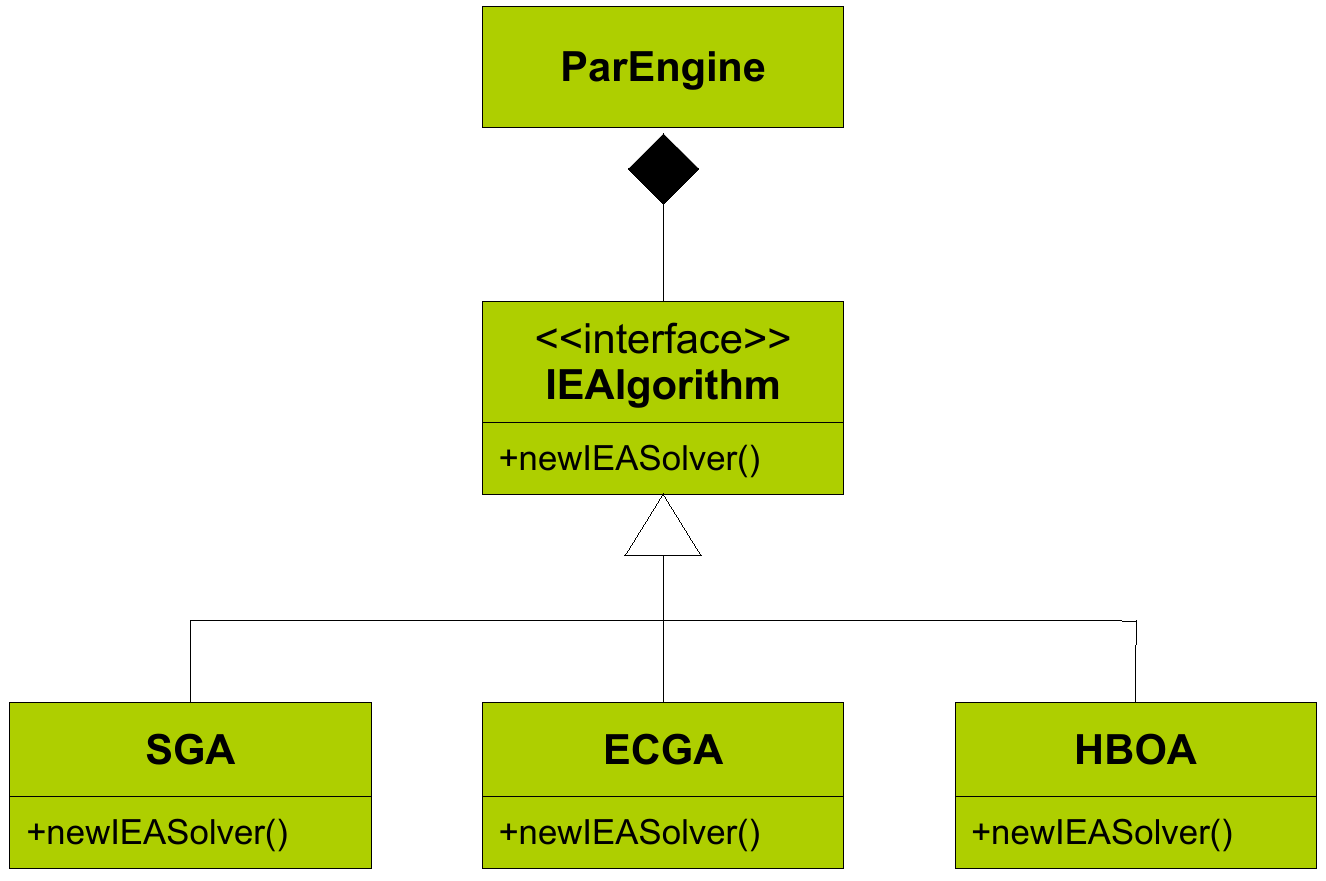}
\caption{The P-EAJava uses the design pattern strategy \citep{Gamma:95} to decouple as much as possible the implementation of a search algorithm from the remaining parameter-less strategy.}
\label{fig:IEAlgUML}
\end{figure}

The P-EAJava also uses the design pattern strategy \citep{Gamma:95} to decouple as much as possible the implementation of a search algorithm from the remaining parameter-less strategy (see Figure \ref{fig:IEAlgUML}). However, as expected, to create a new parameter-less algorithm is not as simple as plugging in a new problem. 

Let us look at the example of the parameter-less SGA to better understand all the necessary steps.
 Note that, the reading of the following instructions is best complemented with a good analysis of the corresponding source code.

\begin{enumerate}
\item Define the class \textit{SGA} that implements the interface IEAlgorithm. Here is the important part of this implementation:

\

\

\vspace{-1.5cm}

\begin{figure}[h!]
\centering
\includegraphics[width=.6\textwidth]{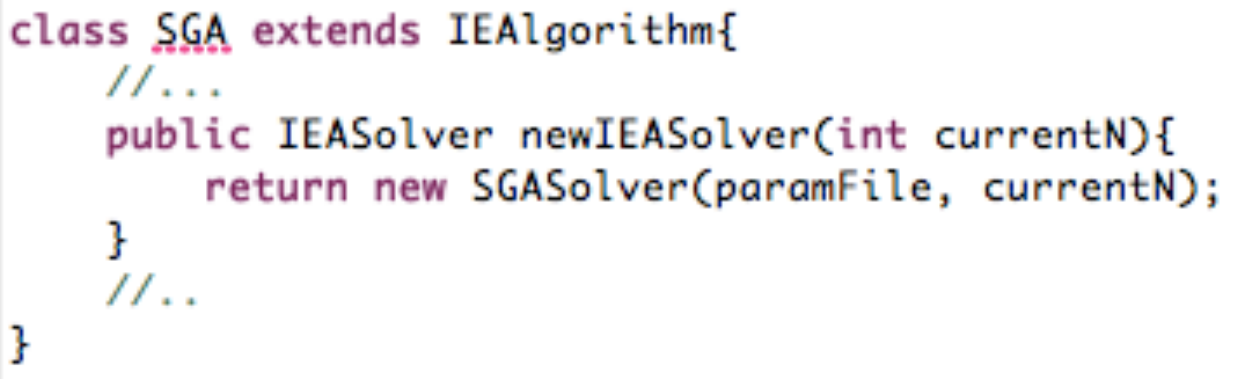}
\end{figure}

\vspace{-.5cm}
Note that, the function \textit{newIEASolver}(int) has  the return type IEASolver which is itself another interface. The class \textit{SGA} is included in the file \textit{src/com/z\_PEA/SGA.java}.

\item Define the class \textit{SGASolver} that implements the interface IEASolver. This class includes the all-important function \textit{nextGeneration()}, responsible for implementing the evolution of the population in each generation. Here is the main part of the implementation:

\newpage
\begin{figure}[h!]
\centering
\includegraphics[width=0.8\textwidth]{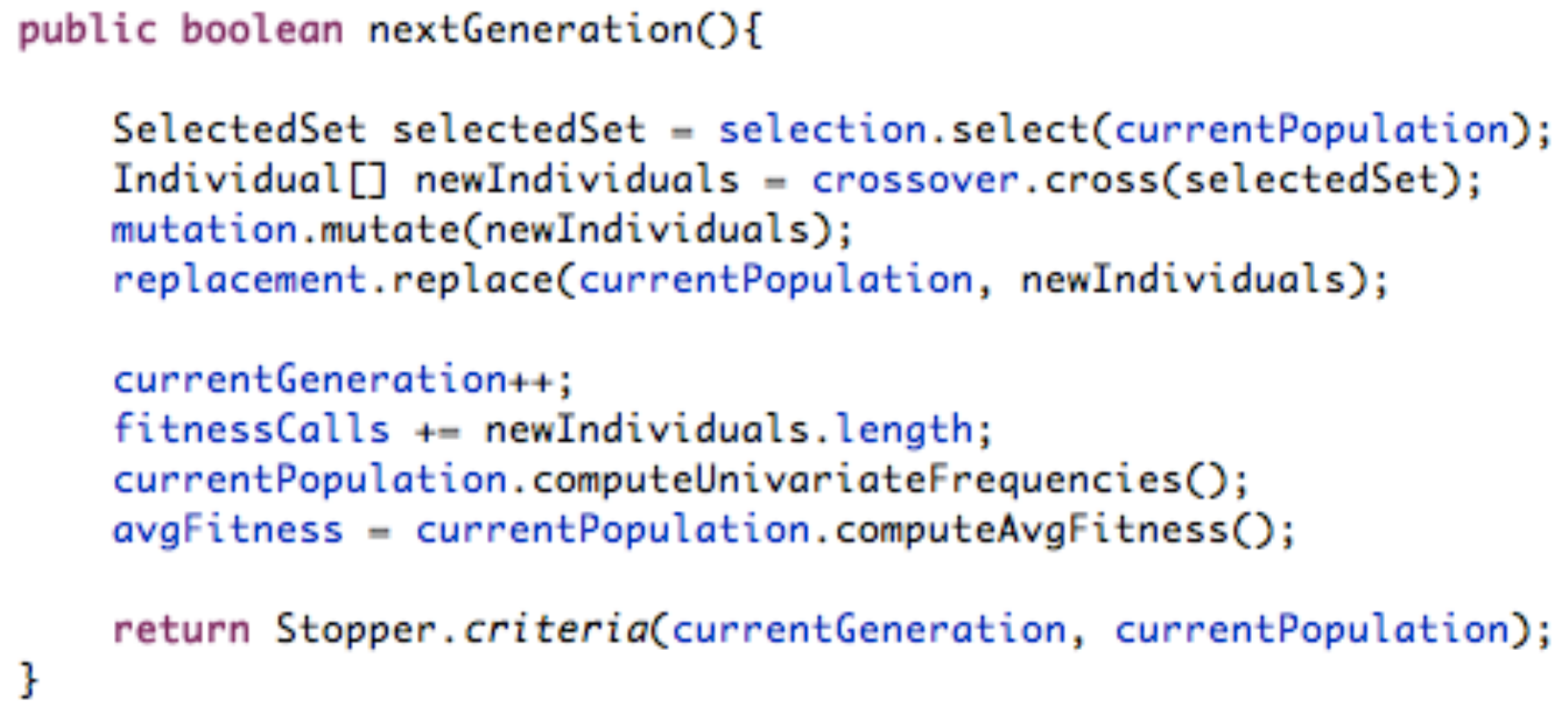}
\end{figure}

\vspace{-.3cm}
Note that, the function must return a boolean as the result of evaluating the stop criteria. The class \textit{SGASolver} is included in the file \textit{src/com/SGA/SGASolver.java}.

\item To define the new algorithm option, add the line
\vspace{-.2cm}
\begin{figure}[h!]
\centering
\includegraphics[width=.45\textwidth]{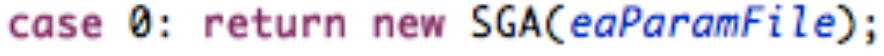}
\end{figure}

\vspace{-.5cm}
 to the \textbf{switch} command in line 241 of the file \textit{src/com/z\_PEA/ParParameter.java}. 
 
\item Validate the new problem option value -- 0 -- by adding the case \textit{eAlg} == 0 to the conditional in line 139 of the same \textit{ParParameter.java} file.
\end{enumerate}

In order to achieve complete integration in the P-EAJava, the parameter-less SGA must also use the classes provided by the framework that define, for instance, the stop criteria of the algorithm. Here is a brief description of those classes:
\begin{description}
\item[Population] Contains the array of individuals that are the population. It provides all the functionalities necessary for operating that set of individuals. In particular, it is responsible for computing its own average fitness, which is a crucial information for the parameter-less strategy.

\item[RandomPopulation] Subclass of the class Population. The constructor of this class is responsible for generating the initial random population.   

\item[SelectedSet] Subclass of the class Population. For some algorithms, depending on the selection operator, the size of this set is different from the population size.

\item[Individual] Contains the string of zeros and ones that represents an individual. It provides all the functionalities necessary for operating that string. In particular, it is responsible for computing its own fitness, according to the problem at hand.

\item[Stopper]  Contains all the stop criteria used by the parameter-less algorithms. These criteria are integrated in a single function called \textit{criteria}(...) which in turn must be returned by the \textit{nextGeneration()} function (see step 2 of the SGA instructions). The criteria options can be changed in the file \textit{PEAParameters.txt}

\end{description}

\section*{Acknowledgements}
The current version of the P-EAJava is one of the by-products of the activities performed by the PhD fellow, José C. Pereira within the doctoral program entitled \textit{Search and Optimization with Adaptive Selection of Evolutionary Algorithms}. The work program is supported by the Portuguese Foundation for Science and Technology with the doctoral scholarship SFRH/BD/78382/2011 and with the research project PTDC/EEI-AUT/2844/2012.



\end{document}